\documentclass[final,5p,times,twocolumn]{elsarticle}

\usepackage{graphicx}

\usepackage{amssymb}

\journal{Journal of Physics and Chemistry of Solids}

\begin{document}

\begin{frontmatter}



\title{Fermi surfaces and quasi-particle band dispersions of the iron pnictides superconductor KFe$_2$As$_2$ observed
by angle-resolved photoemission spectroscopy}


\author[1,2]{T. Yoshida\corref{cor1}}\ead{yoshida@wyvern.phys.s.u-tokyo.ac.jp}
\author[1]{I. Nishi}
\author[1,2]{A. Fujimori}
\author[3]{M. Yi}
\author[3]{R. G. Moore}
\author[3]{D.-H. Lu}
\author[3]{Z.-X. Shen}
\author[2,4]{K. Kihou}
\author[2,4]{P. M. Shirage}
\author[2,4]{H. Kito}
\author[2,4]{C. H. Lee}
\author[2,4]{A. Iyo}
\author[2,4]{H. Eisaki}
\author[2,5]{H. Harima}

\address[1]{Department of Physics, University of Tokyo, Tokyo 113-0033, Japan}
\address[2]{JST, Transformative Research-Project on Iron Pnictides (TRIP), Chiyoda-ku, Tokyo 102-0075, Japan}
\address[3]{Department of Applied Physics and Stanford Synchrotron Radiation Laboratory,
Stanford University, Stanford, CA94305, USA}
\address[4]{National Institute of Advanced Industrial Science and Technology (AIST), Tsukuba 305-8562, Japan}
\address[5]{Department of Physics, Kobe University, Kobe, Hyogo 657-8561, Japan}

\cortext[cor1]{Corresponding author.}

\begin{abstract}
We have performed an angle-resolved photoemission study of the
iron pnictide superconductor KFe$_2$As$_2$ with $T_c\sim$ 4 K.
Most of the observed Fermi surfaces show almost two-dimensional
shapes, while one of the quasi-particle bands near the Fermi level
has a strong dispersion along the $k_z$ direction, consistent with
the result of a band-structure calculation. However, hole Fermi
surfaces $\alpha$ and $\zeta$ are smaller than those predicted by
the calculation while other Fermi surfaces are larger. These
observations are consistent with the result of a de Haas-van
Alphen study and a theoretical prediction on inter-band
scattering, possibly indicating many body effects on the
electronic structure.
\end{abstract}


\end{frontmatter}



\section{Introduction}

\begin{figure*}[htbp]
 \includegraphics[width=18cm]{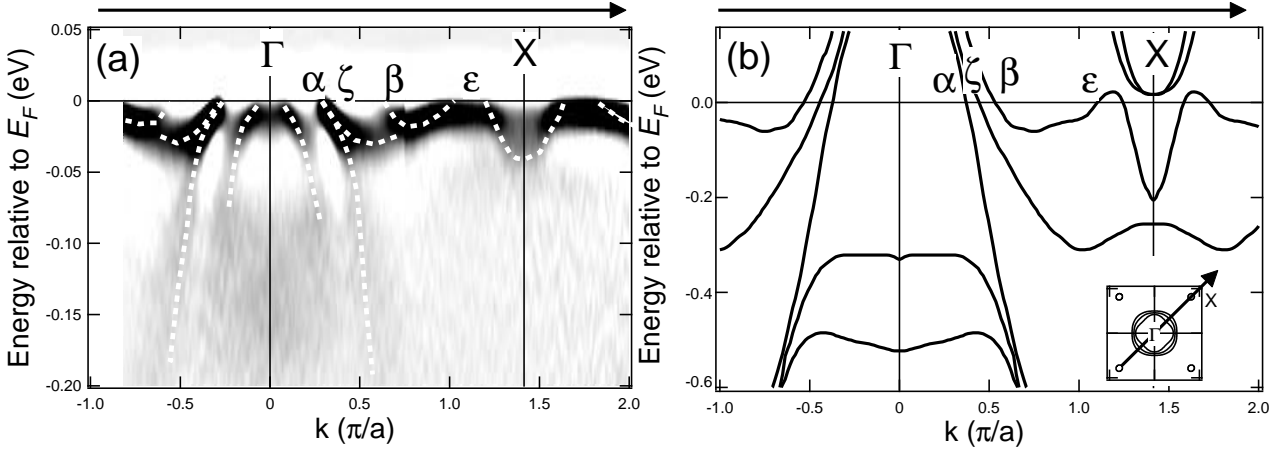}
 \caption{Band dispersion in KFe$_2$As$_2$.
(a) Second derivative plot of the energy distribution curves along
the $\Gamma$-$X$ direction. Dotted lines are guide to the eye. (b)
Band dispersions predicted by band-structure calculation.
}\label{Ek}
 \end{figure*}

The discovery of the iron pnictides superconductors
\cite{Kamihara} has provided a new direction in the studies of
high-$T_c$ superconductors. Similar to the high-$T_c$ cuprates
superconductors, introducing charge carriers (electrons or holes)
into the parent antiferromagnetic material causes
superconductivity. Thus, comparison of the phase diagram of the
iron pnictides with that of the high-$T_c$ cuprates is necessary
for understanding the mechanism of the superconductivity in the
pnictides and the cuprates. In the hole-doped system
Ba$_{1-x}$K$_x$Fe$_2$As$_2$ (Ba122), when the FeAs plane has as
many holes as $\geq$ 0.2 per Fe atom, $T_c$ does not disappear,
different from the high-$T_c$ cuprate. The end member of the Ba122
system KFe$_2$As$_2$, which has 0.5 hole per Fe site, shows
$T_c\sim$ 4 K \cite{Terashima}. Interestingly, $^{75}$As nuclear
quadrupole resonance (NQR) and specific heat studies of
KFe$_2$As$_2$ have suggested the existence of multiple nodal
superconducting gaps rather than a full gap. Also, the electronic
specific heat coefficient $\gamma$ is as large as $\sim$70
mJ/K$^2$mol \cite{Fukazawa}, indicating electron mass
renormalization by electron correlation. In order to study the
relationship between the electron correlation strength and
superconductivity, it has been desired to reveal electron energy
band dispersions and Fermi surfaces (FSs) in KFe$_2$As$_2$. A
previous angle-resolved photoemission study (ARPES) has revealed
that there are hole Fermi surfaces ($\alpha$ and $\beta$) around
the zone center while the electron pockets around the zone corner
disappear and change to small hole pockets ($\epsilon$) due to the
heavy hole doping \cite{Sato}. However, three hole Fermi surfaces
around the zone center predicted by the band-structure calculation
has not been resolved. Recently, a de Haas-van Alphen (dHvA) study
has revealed that the $\zeta$ sheet which has a similar size to
the $\alpha$ sheet exists around the zone center \cite{dHvA}.
Furthermore, the dHvA study has pointed out the shrinkage of the
$\alpha$ and $\zeta$ sheets and the enhancement of electron mass
compared to those predicted by the band-structure calculation. In
the present study, in order to reveal the shapes of the FSs more
precisely in the three-dimensional momentum space, we have
performed an ARPES study of KFe$_2$As$_2$ using high-quality
single crystals and various photon energies.

\begin{figure}[htbp]
 \includegraphics[width=7.5cm]{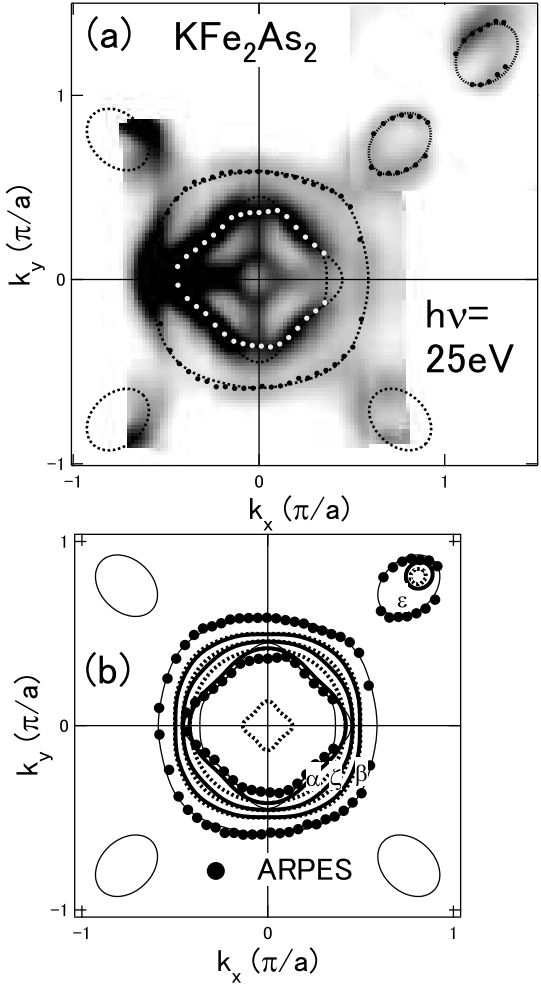}
 \caption{Fermi surfaces of KFe$_2$As$_2$ obtained by
ARPES. (a) Spectral weight at $E_F$ in the $k_x$-$k_y$ plane.
Fermi momenta $k_F$ (filled circles) were obtained by peak
positions of the momentum distribution curves. (b) Comparison of
the FSs obtained by ARPES (filled circles) with those predicted by
band-structure calculation. Results of band-structure calculation
are shown by thick solid lines ($k_z$= 0) and thick dotted lines
($k_z$= $2\pi/c$ corresponding to the Z point).}\label{FS2D}
 \end{figure}

\section{Experiment and Calculation}
ARPES measurements were performed at beamline 5-4 of Stanford
Synchrotron Radiation Laboratory (SSRL) with a normal incidence
monochromator and a Scienta SES-R4000 electron analyzer. The
typical energy and angular resolutions used for the present
measurements were about 10 meV and 0.3 degree, respectively.
Single crystals of KFe$_2$As$_2$ were grown from K flux. Samples
were cleaved \textit{in situ} and measured at a temperature of 15
K in a pressure better than $5 \times 10^{-11}$ Torr. We have
performed the measurements at photon energies from $h\nu$=14 to 33
eV. The in-plane ($k_x$, $k_y$) and out-of-plane ($k_z$) momentum
are expressed in units of $\pi/a$ and $2\pi/c$, respectively,
where $a= 3.864$ \textrm{\AA} and $c=13.87$ \textrm{\AA}. The
electronic band structure of KFe$_2$As$_2$ was calculated within
the local density approximation (LDA) by using the full potential
LAPW (FLAPW) method. We used the program codes TSPACE
\cite{Yanase} and KANSAI-06. The experimental crystal structure
\cite{Rozsa} including the atomic position $z_\mathrm{As}$ of As
was used for the calculation.

\section{Results and Discussion}
Band dispersions taken with $h\nu$=25 eV are shown in Fig.
\ref{Ek}(a). The cut is along the diagonal of the two-dimensional
Brillouin zone and the image has been obtained by the second
derivative of the energy distribution curves. All the energy bands
giving rise to the $\alpha$, $\zeta$, $\beta$, and $\epsilon$ FS
sheets predicted by the calculation [Fig.\ref{Ek}(b)] are
observed. Particularly, we have clearly observed the $\zeta$ band
close to the $\alpha$ band near $E_F$, consistent with the dHvA
result \cite{dHvA}. In addition to these bands, we find that
another hole-like band exists near the zone center. Since this is
not predicted by the bulk band-structure calculation, this band
may be due to surface states. By comparing the ARPES band
dispersions with the band-structure calculation shown in
Fig.\ref{Ek} (b), the enhancement of electron mass of the $\alpha$
band is found to be $m^*/m_b\sim$ 3. On the other hand, the
dispersion of the $\zeta$ band strongly deviates from the
band-structure calculation, indicating orbital dependent mass
renormalization.

Figure \ref{FS2D} (a) shows Fermi surfaces in the $k_x$-$k_y$
plane of KFe$_2$As$_2$ obtained by ARPES. The overall Fermi
surface shapes are nearly consistent with those observed in the
previous study \cite{Sato}. However, several new observations
should be remarked. One is the observation of the $\zeta$ sheet
which has a similar size as the $\alpha$ sheet as already pointed
out above. Fermi momenta $k_F$'s obtained by the MDC peak
positions show different values between the $k_x$ and the $k_y$
direction. This can be explained by existence of the two Fermi
surfaces ($\alpha$ and $\zeta$) with different matrix element
effects. According to the dHvA result \cite{dHvA}, the larger hole
Fermi surface is ascribed to the $\zeta$ Fermi surface. Another
point is that a small Fermi surface around the zone center has
been observed. Since this FS is not predicted by the
band-structure calculation and has nearly two-dimensional
dispersions as indicated below, it can be ascribed to surface
states.

\begin{figure*}[htbp]
\begin{center}
\includegraphics[width=15cm]{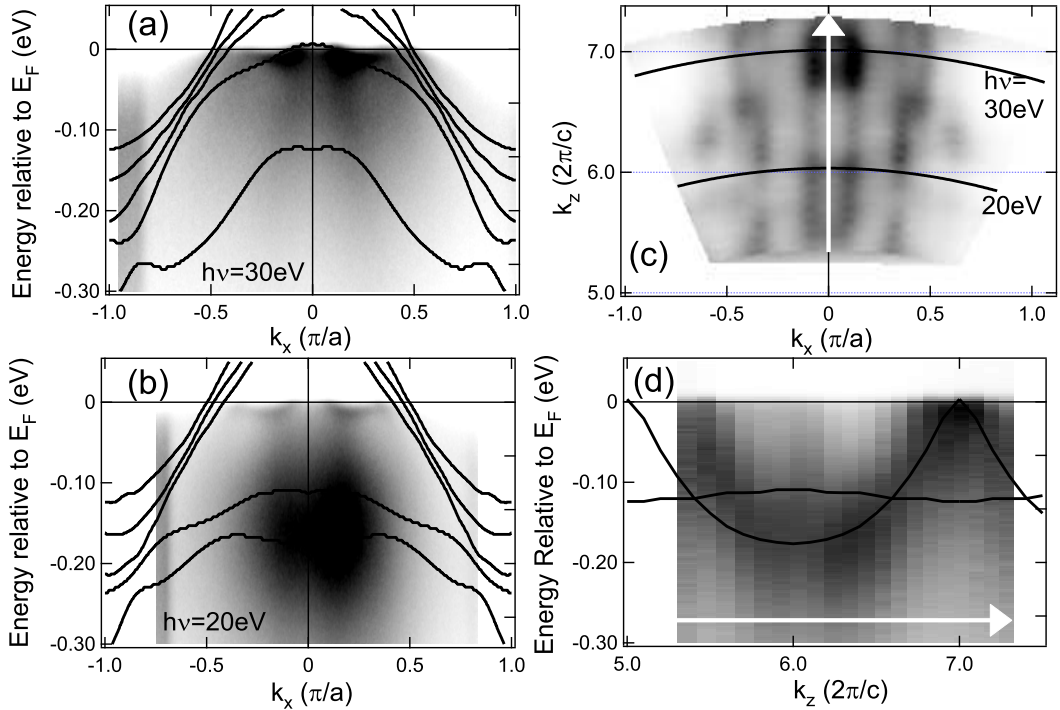}
\caption{Three-dimensional electronic structure of KFe$_2$As$_2$.
(a)(b) ARPES spectra along the $k_x$ direction with $k_y$= 0 taken
at $h\nu$= 20 eV and 30 eV. Thick lines are band dispersions
predicted by the band-structure calculation for $k_z$= $2\pi/c$
[(a)] and $k_z$= 0 [(b)]. (c) Spectral weight mapping at $E_F$ in
the $k_x$-$k_z$ plane. (d) Normal emission spectra corresponding
to the arrow in panel (c).}\label{FS3D}
\end{center}
\end{figure*}

 Here, we compare the FSs obtained by ARPES with those
predicted by the band-structure calculation in Fig.\ref{FS2D} (b).
The Fermi surface areas are estimated to be 10.1, 11.8, 28.5 and
2.1 \% of the Brillouin zone for $\alpha$, $\zeta$, $\beta$, and
$\epsilon$, respectively, in good agreement with the results of
dHvA \cite{dHvA}. Note that the $\alpha$ and $\zeta$ Fermi
surfaces are smaller than those in the band-structure calculation,
while the $\beta$ and $\epsilon$ FSs are larger. This trend may be
explained by a theoretical prediction based on inter-band
interaction \cite{Ortenzi}.

Since the Ba122 system has three dimensional FSs which are
predicted by band-structure calculations \cite{Singh} and
confirmed by ARPES \cite{Malaeb,Vilmercati}, we have investigated
the electronic band dispersions in the $k_z$ direction by changing
the excitation photon energy. Figuer \ref{FS3D}(c) shows spectral
weight mapping at $E_F$ in the $k_x$-$k_z$ plane. $k_z$ has been
is determined by assuming an inner potential $V_0$=13eV. ARPES
spectra near the zone center taken at $h\nu$= 20 eV and 30 eV,
corresponding to the cuts in Fig. \ref{FS3D}(c), are shown in Fig.
\ref{FS3D} (a) and Fig. \ref{FS3D}(b), respectively. From Fig.
\ref{FS3D}(c), one can see that the observed Fermi surfaces are
nearly two-dimensional, consistent with the band-structure
calculation. Note that the FS of the surface state near the zone
center has also two-dimensional character. From Figs.
\ref{FS3D}(a) and \ref{FS3D}(b), one can see that the structure
$\sim$ 0.2 eV below $E_F$ observed for h$\nu$=20 eV moves towards
$E_F$ for h$\nu$=30 eV, indicating that the this band is
dispersive along the $k_z$ direction. In fact, normal emission
spectra [Fig.\ref{FS3D} (d)] clearly illustrate a parabolic
dispersion along the $k_z$ direction and slightly crosses $E_F$ in
the vicinity of the $Z$ point. From the band-structure
calculation, this band can be identified as the $d_{z^2}$ band.
With a mass enhancement factor $m^*/m_b\sim$ 3, this band is also
well explained by the band-structure calculation.

\section{Conclusion}
In summary, we have observed FSs in KFe$_2$As$_2$ in
three-dimensional momentum space by using ARPES. All FSs except
for the surface states are qualitatively consistent with the
band-structure calculation. The sizes of the FSs are
quantitatively consistent with the dHvA observation: the $\alpha$
and $\zeta$ sheets are smaller than those in the band-structure
calculation while the $\beta$ and $\epsilon$ sheets are larger,
which may be attributed to electron correlation effects.

\section*{Acknowledgement}
We are grateful to H. Fukazawa, T. Terashima and M. Kimata for
enlightening discussions. This work was supported by
Transformative Research-Project on Iron Pnictides (TRIP) from the
Japan Science and Technology Agency, and the Japan-China-Korea A3
Foresight Program from the Japan Society for the Promotion of
Science. SSRL is operated by the US DOE Office of Basic Energy
Science Divisions of Chemical Sciences and Material Sciences.



\end{document}